# LOW COMPLEXITY TIME DOMAIN SEMI-BLIND MIMO-OFDM CHANNEL ESTIMATION USING ADAPTIVE BUSSGANG ALGORITHM

Ebrahim Karami and Markku Juntti

Centre for Wireless Communications (CWC), University of Oulu,

P.O. Box 4500, FIN-90014, Oulu, Finland

ABSTRACT

In this paper, a low complexity time domain semi-blind algorithm is proposed to estimate and track the time varying MIMO OFDM channels. First, the proposed least mean squares (LMS) based algorithm is developed for the training mode and then is extended for the blind mode of the operation by combining with the decision direction (DD) or adaptive Bussgang algorithm (ABA) techniques. In the blind mode, because of decision errors, a smaller step size is considered for the LMS algorithm and the channel estimation is run a few times to improve its precision. In each round of the estimation in the blind mode, the step size is decreased to form some kind of annealing. Both DD LMS and ABA LMS techniques are simulated and compared to the full training case and MSE of channel estimation error is considered as comparison criterion. It is shown for $2 \times 4$ DD LMS and for $4 \times 4$ ABA LMS algorithms present near full training case estimation error. Of course in some scenarios the former proposed technique performs better and in other scenarios the latter is better and therefore combine of it can be very interesting in all channel conditions.

## I. INTRODUCTION

In recent years, MIMO channels are introduced to achieve high data rates required by the next generation wireless communication systems [1]. Using multiple-input multiple output (MIMO) channels, when bandwidth is limited, provides much higher spectral efficiency compared to Single-Input Single-Output (SISO), Single-Input Multiple-Output (SIMO), and Multiple-Input Single-Output (MISO) channels. Moreover, when channel is full rank or in other words all propagation coefficients are independent, the diversity gain of MIMO channels is product of transmitter and receiver elements numbers. Therefore, employing MIMO channels not only increases the throughput, but also increases their robustness against fading; thus, makes it efficient for the requirements of the next generation wireless services. Using orthogonal frequency division multiplexing (OFDM) combats the frequency selective fading by converting a wide band channel to a couple of narrowband flat fading channels [2]. To detect the received signal in a time varying wireless channel, use of either equalization or channel estimation is obligatory [3-10]. While use of the channel estimation, designer has more flexibility to use more kinds of the different detection and decoding algorithms than equalization and therefore channel estimation is a more common option.

Estimation of the MIMO OFDM channels can be done in the time domain or frequency domain [11]. The most researches done in the MIMO OFDM channel estimation topic, is in frequency domain where because of especial features of the OFDM signals, implementation of the algorithms is more feasible. But while using the frequency domain channel estimation, information related to the delay spread of the channel paths is ignored and this simplification decreases the performance. Usually to compensate this degradation, a mapping to the time domain and then channel shortening is used. But this process needs, whole of an OFDM symbol is assigned to the training, or in other word all sub-carriers must be available for designer. This requirement not only decreases the effective throughput but also in some scenarios is not feasible because all sub-carriers are not available. Therefore, time domain channel estimation provides more flexibility for the designer and better performance compared to the frequency domain algorithms. On the other hand, when channel undergoes time variations, the algorithm should follow channel variations in the blind mode and the extension off the time domain algorithms for the blind mode is strait-forward. Two conventional techniques to extend training based channel estimation and equalization algorithms to the semi-blind mode are decision directed [12-14] and modulation-statistics-based algorithms [15-18] which respectively use hard and soft decision detected data as virtual training. Bussgang technique which uses a non-linear function of soft detected data is conventionally used in equalization and its application as a channel estimator is novel [19]. The optimum non-linear function used in Bussgang algorithms is dependent to the SNR and other channel parameters, therefore its adaptation provide improvement in the performance [11]. In this paper, LMS algorithm is developed as a semi-blind time domain estimation algorithm for time varying MIMO channel and to adapt non-linear function used by Bussgang algorithm, two parameters in it are considered to be adapted.

The rest of this paper is organized as follows. In Section II, models used for signal transmission and channel are introduced. In Section III, the proposed algorithm is derived. In Section IV simulation results of the proposed receiver are presented. Concluding remarks are presented in Section V.

## II. SYSTEM MODEL

The block diagram of the transmitter in a spatial multiplexed MIMO-OFDM system with $M$ antennas is shown in Fig. 1 [2].

The main input block is converted to a couple of parallel sub-streams using a serial to parallel converter. Then each sub-stream is converted to $M$ OFDM symbols. Finally, after inserting cyclic prefix with minimum length equal to the delay spread of the channel, all $M$ sub-blocks are transmitted separately via transmitters.

In the receiver side, linear combinations of all transmitted sub-blocks are distorted by time-varying Rayleigh or Ricean fading, and the inter symbol interference (ISI) are observed under the additive white Gaussian noise.

Frequency selective fading MIMO channel with time varying Rayleigh distribution and exponentially decaying paths is assumed. The input/output equations can be summarized as

$$r_k = \sum_{p=0}^{L_p-1} H_{k,p} s_{k-p} + w_k, \quad k=0, 1, \ldots, L+L_{cp}-1 \quad (1)$$

where $L_{cp}$ and $L_p$ are the length of inserted cyclic prefix and the number of resolvable paths, and $r_k$ is the received vector, $H_{k,p}$ is the channel matrix, and $s_k$ is the transmitted symbol all in time index $k$, and $w_k$ is the vector with i.i.d. AWGN elements with variance $\sigma_w^2$. Of course usually channel matrices have slow variation on an OFDM block and therefore sub-script $k$ can be omitted. This is very important in OFDM systems. Because of ICI is produced by variation of the channel coefficients over an OFDM block.

At the receiver side as shown in Fig .2, in each antenna, after removing cyclic prefix, each sub-block is fed to a FFT. If $L_{cp}$ is greater than $L_p$, the net input/output equation in time domain will be as,

$$y_l = \tilde{H}_l x_l + z_l, \quad l=0,1,\ldots, L, \quad (2)$$

where $x_l$ and $y_l$ are the input vector in the bank of IFFT in transmitter and the output of the FFT bank in receiver and $z_k$ is the equivalent noise vector in the output of the FFT at receiver and is FFT of noise vector with the same variance i.e. $\sigma_w^2$ and $\tilde{H}_l$ is frequency domain channel matrix for sub-band $l$ and defined as

$$\tilde{H}_l = \sum_{p=0}^{L_p-1} H_p \exp\left(\frac{j2\pi pl}{N_c}\right). \quad (3)$$

Finally, the problem pf semi-blind MIMO OFDM channel estimation can be summarized as estimation of $H_p$ s matrixes when all $y_l$ s and some $x_l$ s are known.

### III. THE PROPOSED ALGORITHM

The block diagram of the receiver is shown in Fig. 2. The basis of the proposed technique is based on the LMS algorithm, where the following cost function is minimized.

$$C(H_0, H_1, \ldots, H_{L_p-2}, H_{L_p-1}) = E\left\langle \left\| y_k - \left( \sum_{p=0}^{L_p-1} H_p \exp\left(-\frac{j2\pi pk}{N_c}\right) \right) x_k \right\|^2 \right\rangle \quad (4)$$

where $x_k$ and $y_k$ are transmitter and receiver vectors respectively, $H_p$ s are $p$th channel matrix coefficients, $L_p$ is the length of the channel, $N_c$ is the number of sub-carriers in OFDM symbols, and $E\langle . \rangle$ and $\|.\|$ and expectation and norm operators. $C(H_0, H_1, \ldots, H_{L_p-2}, H_{L_p-1})$ is minimized iteratively with respect to the channel matrixes iteratively and by the following LMS type equation.

$$H_m^{(n)} = H_m^{(n-1)} + \mu_{m,n} \left[ y_k - \sum_{p=0}^{L_p-1} H_p^{(n-1)} \exp\left(\frac{j2\pi pk}{N_c}\right) x_k \right] x_k^H \exp\left(\frac{j2\pi pk}{N_c}\right) \quad (5)$$

where $H_m^{(n)}$ is $m$th channel matrix coefficient estimated in the $n$th time and $\mu_{m,n}$ is the step size.

In the blind mode of the operation, (5) is extended by DD and ABA techniques as follows,

$$H_m^{(n)} = H_m^{(n-1)} + \mu_{m,n} \left[ y_k - \sum_{p=0}^{L_p-1} H_p^{(n-1)} \exp\left(\frac{j2\pi pk}{N_c}\right) g(\tilde{x}_k)^H \right] g(\tilde{x}_k)^{HH} \exp\left(\frac{j2\pi pk}{N_c}\right), \quad (6)$$

where $\tilde{x}_k$ is soft estimation of $x_k$ and $g(.)$ is a nonlinear function. When BPSK signaling is used, the best selection of the $g(.)$ for the DD algorithm is signum function and for the ABA $g(\tilde{x}_k) = \alpha_n \tanh(\beta_n \tilde{x}_k)$, and the values for the $\alpha_n$ and $\beta_n$ are adaptive parameters optimized by the following iterative equations.

$$\alpha_n = \alpha_{n-1} + \mu_\alpha \operatorname{sgn}\left[ \sum_{i=1}^{M} \tanh(\beta_{n-1}\tilde{x}_k) \left( \hat{y}_k^i - \alpha_{n-1} \tanh(\beta_{n-1}\tilde{x}_k) \right) \right], \quad (7)$$

$$\beta_n = \beta_{n-1} + \mu_\beta \alpha_{n-1} \sum_{i=1}^{M} \tilde{x}_k \operatorname{sec} h(\beta_{n-1}\tilde{x}_k)^2 \times (\tilde{x}_k - \alpha_{n-1} \tanh(\beta_{n-1}\tilde{x}_k)). \quad (8)$$

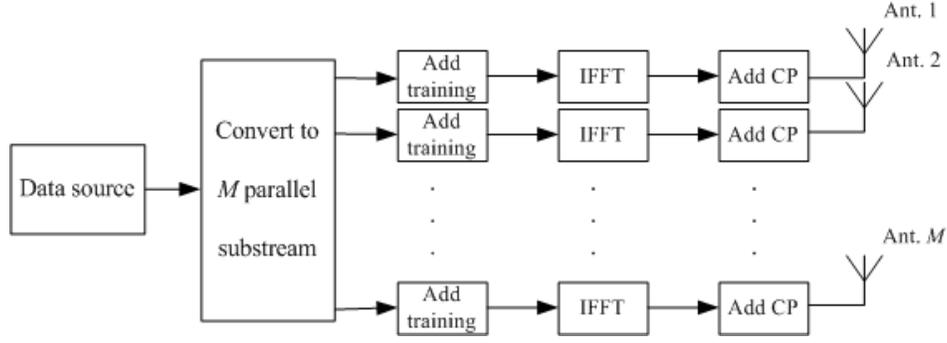

Fig. 1. Block diagram of the transmitter.

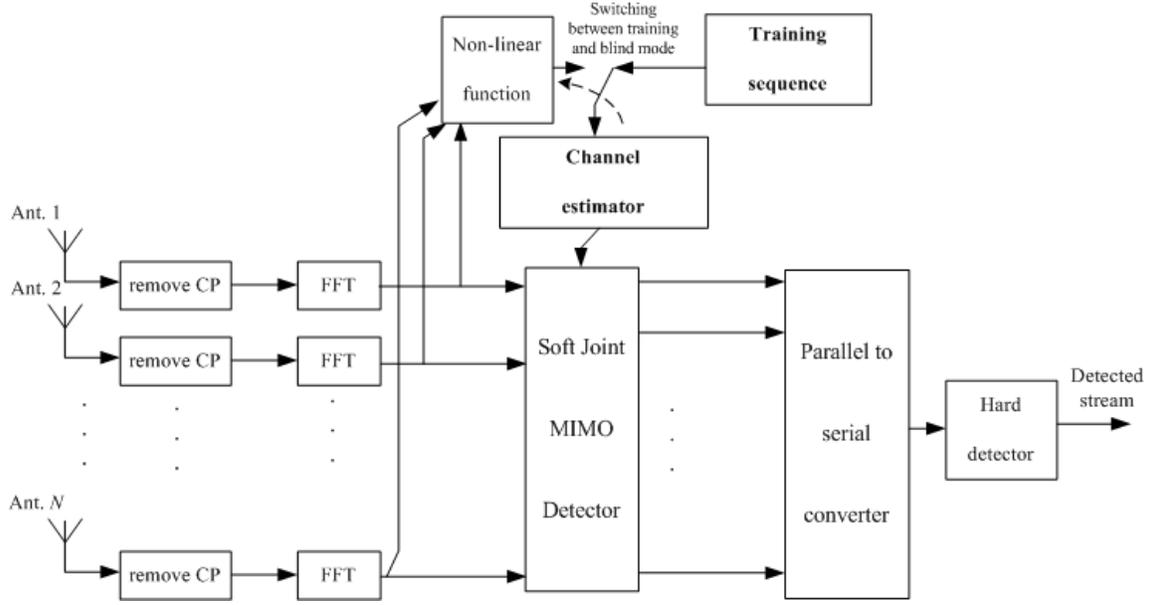

Fig. 2. Block diagram of the proposed receiver.

When using a hard detector $\tilde{x}_k$ can be calculated from hard detected data vector by the following equation

$$\tilde{x}_k = \left( \sum_{p=0}^{L_p-1} H_p^{(n-1)} \exp\left( \frac{j2\pi pk}{N_c} \right) \right)^H y_k - F_k \hat{x}_k \qquad (9)$$

where $\hat{x}$ is the hard detected of the $x_k$ and $F_k$ is defined as

$$F_k = \left[ \left( \sum_{p=0}^{L_p-1} H_p^{(n-1)} \exp\left( \frac{j2\pi pk}{N_c} \right) \right)^H \right. \\ \left. \times \left( \sum_{p=0}^{L_p-1} H_p^{(n-1)} \exp\left( \frac{j2\pi pk}{N_c} \right) \right) \right]_{ZD} \qquad (10)$$

where $[.]_{ZD}$ is defined as a operator which puts zeros on the diagonal elements of the matrixes. Therefore the algorithm can be summarized as:

Step 1. Initialization of estimated channel matrixes to zero, i.e. $H_p^{(0)}$, $p = 0,1,...,L_p-2, L_p-1$ and $\alpha_0$ and $\beta_0 = 1$ for ABA case.

Step 2. Run the algorithm in the blind mode for all subcarriers assigned to the training using equation (5).

Step 3. Run the algorithm in the blind mode as follows,
  A. Hard detection of transmitted vector by any interested algorithm.
  B. Calculation of the soft transmitted vector by equations (9) and (10).
  C. Updating estimated channel matrixes by equation (3) and adaptation of $\alpha_n$ and $\beta_n$ by equations (9) and (10) in the ABA algorithm.

Step 3 which is for the blind mode is run a few times to get the better results. Usually 3-5 iterations is enough to get the best result.

## IV. SIMULATION RESULTS

Simulation results are presented in the two parts. In the first part, the proposed and reference algorithms are simulated for the un-coded signals and a part of each OFDM symbol is dedicated to the training and all algorithms work blind on the other part of each OFDM symbol. But in the second part of the simulations, first OFDM symbol is dedicated to the training and other OFDM symbols is dedicated to the consecutive turbo-coded block of the data.

### A. Un-coded system with both training and data in each OFDM symbol

First, the proposed semi-blind techniques are simulated and compared to the full training and full known cases for the un-coded signals. Simulations are done for 2×4 and 4×4 MIMO-OFDM channels and are summarized in the Figs. 3-6. 512 sub-carriers are assumed and ¼ of first of them, i.e. 128 first sub-carriers are allocated as training and other is for data. Simulations are done for outdoor Winner channel model for fast mobile speed 120km/h. Of course because of running operating independently over OFDM symbols the speed of the mobile user has not noticeable effect on the performance. BER and MSE of channel estimation are considered as comparison criteria.

Figs. 3 and 4 present the comparison of both semi-blind proposed algorithms with full training case considering MSE of the channel estimation as the comparison criterion. These simulations are run for Eb/N0=0dB and Eb/N0=30dB versus the number of running over OFDM symbol. We can see in all cases and especially for 2×4 MIMO OFDM channel, ABA LMS presents the best performance very close to the full training case which is lower bound. Also, in all cases 3 run of the algorithms (3 iterations) presents the minimum MSE and therefore in the next part of the simulations, it is considered as a reference to comparison. Notice using iteration is not conventional in single carrier channel estimation and equalization but here we can get gain using it because while applying LMS as sub-carrier by sub-carrier the effect of first sub-carriers is gradually decreased whereas all sub-carriers should have the same effect. By iteratively applying the algorithm we can force all sub-carriers to have the same effect in the algorithm.

Figs. 5 and 6 present BER comparison of DD LMS and ABA LMS techniques with non-blind LMS and full known channel cases. In all cases and all ranges of the Eb/N0 proposed semi-blind algorithms outperforms non-blind case completely and present much lower BER floors.

When algorithms are run only 1 time, DD LMS presents always the best performance. But when algorithms are run 3 times and in 2×4 channel ABA LMS presents better performance which is very close to full-training case which is an unreachable lower bound.

### B. Coded system with first symbol as training and other ones as data

Simulations results for the coded MIMO OFDM systems are presented in Fig. 7-9. In this case, first symbol of each 20 OFDM symbols is dedicated to the training. Here we also study the tracking behavior of the algorithm under time variations and so semi-blind time domain least square (LS) algorithm [8] has been considered as a reference to compare. Notice in the first part of the simulation, because of too short length of the training, LS algorithm is encountered to inversion of an incomplete matrix and consequently it fails to work, even in the training mode. Each data OFDM symbol is considered as 1 turbo coded block. A rate ½ turbo code with QPSK signaling and constraint length 3 is considered. Channel model is the same as last simulation to show the capability of the algorithm to track fast channel variations.

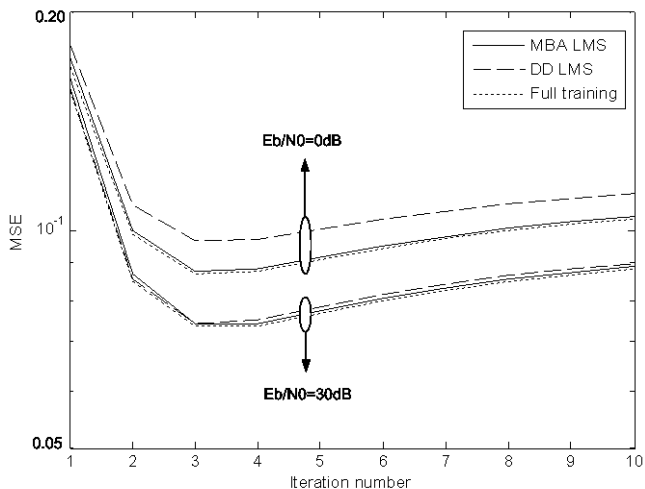

Fig. 3. Comparison of the MSE of the channel estimation error for the 2×4 MIMO OFDM channel.

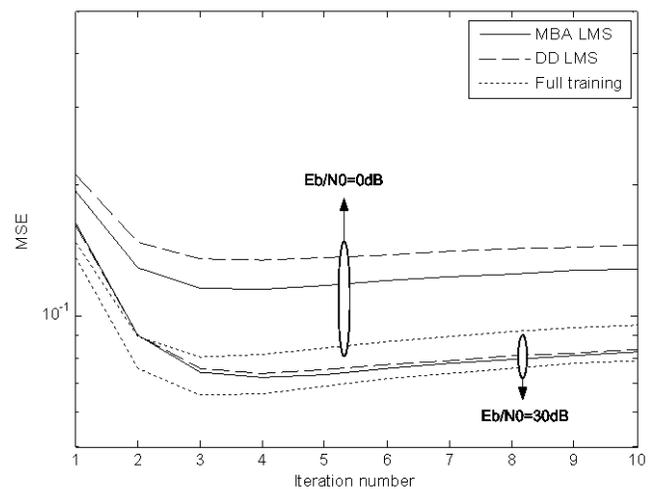

Fig. 4. Comparison of the MSE of the channel estimation error for the 4×4 MIMO OFDM channel.

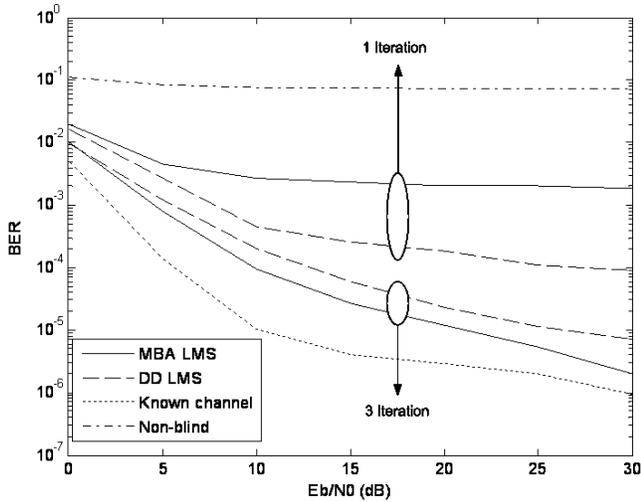

Fig. 5. BER comparison for the 2×4 MIMO OFDM channel (un-coded signals).

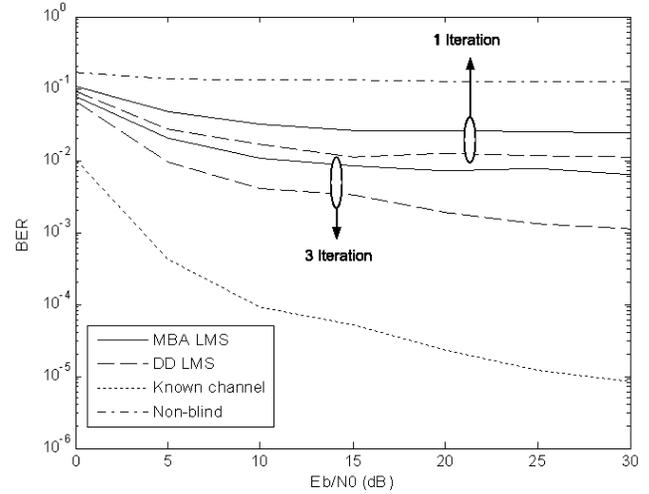

Fig. 6. BER comparison for the 4×4 MIMO OFDM channel (un-coded signals).

The LMS algorithm is compared to the LS algorithm for two cases i.e. first case when all 512 sub-carriers all available and second when only 300 sub-carriers are available and other ones are nulled. Fig. 7 presents the performance of the two algorithms after working in the training mode i.e. after the first OFDM symbol. It can be seen that in the training mode when all sub-carriers are available, LS provides a bit better performance but when a part of OFDM band is available, performance of LMS is much better than LS.

Figs. 8 and 9 show the performance of two algorithms in the tracking mode. Tracking behaviors of the LMS and LS algorithms for data symbols is shown in Fig. 8. In this figure x axis is index of data OFDM symbols which is from 2 to 20. It can be seen that in the both cases, LS algorithm can not track symbol by symbol channel variations but LMS is successful to track channel variations and MSE of the channel estimation is fixed and even with a small negative slope in the DD mode.

And BER results for turbo coded signals is shown is Fig. 9 and you can see in both cases LS algorithm give a high error floor which is due to its inefficiency to track channel variations but LMS provides acceptable BER even when a part of bandwidth is available. In summary, in training mode or when channel is invariant, LMS outperform LS when a part of OFDM symbol is available and in the tracking mode or when channel is time variant in both cases LMS outperforms LS and provides much better performance.

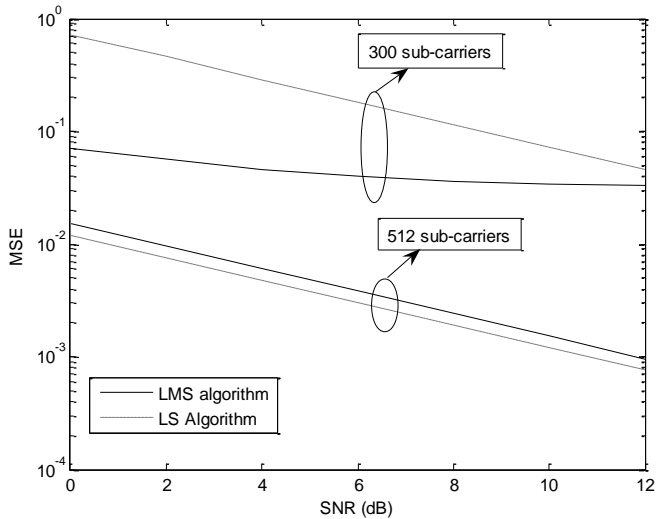

Fig. 7. MSE of the channel estimation error after first OFDM symbol (training symbol).

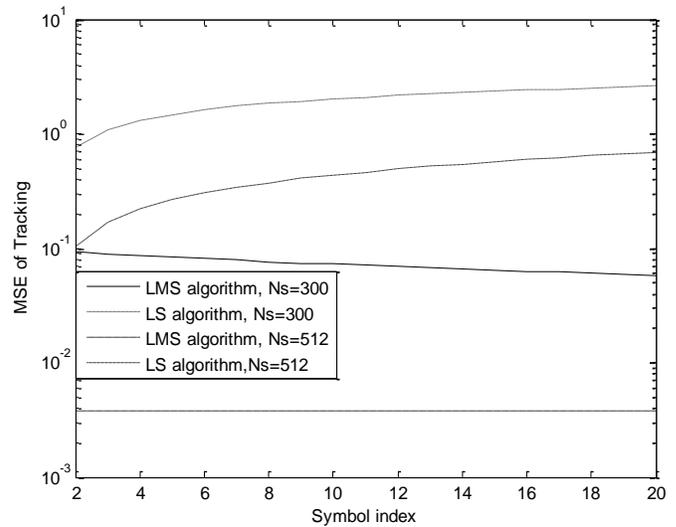

Fig. 8. Tracking behavior of the LS and LMS channel estimation error.

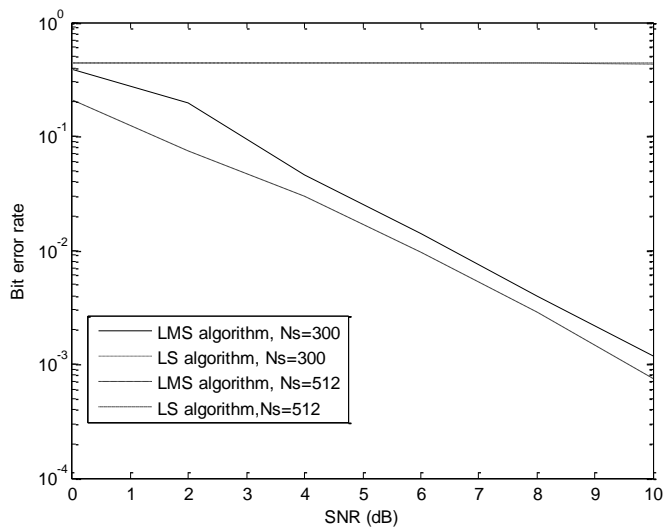

Fig. 9. BER of LMS and LMS channel estimation algorithms for turbo coded signals (turbo coded signals).

## V. CONCLUSION

In this paper, conventional LMS algorithm is extended as a semi-blind MIMO-OFDM channel estimation and tracking algorithm. First the algorithm is derived ow complexity time domain semi-blind algorithm was proposed to estimate and track the time varying MIMO OFDM channels. First, the proposed least mean squares (LMS) based algorithm is developed for the training mode and then is extended for the blind mode of the operation by combining with the decision direction (DD) or adaptive Bussgang algorithm (ABA) techniques. In the blind mode, because of decision errors, a smaller step size is considered for the LMS algorithm and the channel estimation is run a few times to improve its precision. In each round of the estimation in the blind mode, the step size is decreased to form some kind of annealing. Both DD LMS and ABA LMS techniques are simulated and compared to the full training case and MSE of channel estimation error is considered as comparison criterion. It is shown for $2\times4$ DD LMS and for $4\times4$ ABA LMS algorithms present near full training case estimation error. Of course in some scenarios the former proposed technique performs better and in other scenarios the latter is better and therefore combine.

## EREFERENCES